\documentclass{article}

\usepackage{microtype}
\usepackage{graphicx}
\usepackage{subcaption}
\usepackage{booktabs} 

\usepackage{multirow}
\usepackage{amsmath}
\usepackage{textcomp}
\usepackage{hyperref}

\usepackage[preprint]{icml2026}

\usepackage{amsmath}
\usepackage{amssymb}
\usepackage{mathtools}
\usepackage{amsthm}

\usepackage[capitalize,noabbrev]{cleveref}

\theoremstyle{plain}

\theoremstyle{definition}

\theoremstyle{remark}

\usepackage[textsize=tiny]{todonotes}

\icmltitlerunning{Chain-of-Thought and Reinforcement Learning for Target Speaker Automatic Speech Recognition}

\begin{document}

\twocolumn[
  \icmltitle{Thinking in Cocktail Party: Chain-of-Thought and Reinforcement Learning for Target Speaker Automatic Speech Recognition}






  \icmlsetsymbol{equal}{*}

  \begin{icmlauthorlist}
    \icmlauthor{Yiru Zhang}{equal,comp}
    \icmlauthor{Hang Su}{equal,comp}
    \icmlauthor{Lichun Fan}{comp}
    \icmlauthor{Zhenbo Luo}{comp}
    \icmlauthor{Jian Luan}{comp}
  \end{icmlauthorlist}

  \icmlaffiliation{comp}{MiLM Plus, Xiaomi Inc., China. * means equal contribution}

  \icmlcorrespondingauthor{Lichun Fan}{fanlichun1@xiaomi.com}

  \icmlkeywords{Machine Learning, ICML}

  \vskip 0.3in
]



\printAffiliationsAndNotice{}  

\begin{abstract}
Target Speaker Automatic Speech Recognition (TS-ASR) aims to transcribe the speech of a specified target speaker from multi-speaker mixtures in cocktail party scenarios. Recent advancement of Large Audio-Language Models (LALMs) has already brought some new insights to TS-ASR. However, significant room for optimization remains for the TS-ASR task within the LALMs architecture. While Chain of Thoughts (CoT) and Reinforcement Learning (RL) have proven effective in certain speech tasks, TS-ASR, which requires the model to deeply comprehend speech signals, differentiate various speakers, and handle overlapping utterances is particularly well-suited to a reasoning-guided approach. Therefore, we propose a novel framework that incorporates CoT and RL training into TS-ASR for performance improvement. A novel CoT dataset of TS-ASR is constructed, and the TS-ASR model is first trained on regular data and then fine-tuned on CoT data. Finally, the model is further trained with RL using selected data to enhance generalized reasoning capabilities. Experiment results show a significant improvement of TS-ASR performance with CoT and RL training, which demonstrates the effectiveness of the proposed CoT and RL training methods adapted for the TS-ASR task.
\end{abstract}

\section{Introduction}

Large Language Models (LLMs), which originally developed for text understanding and generation, have achieved significant advances by scaling the parameters and the data exponentially~\cite{gpt4, LLaMA}. In the field of speech and audio processing, Large Audio-Language Models (LALMs) have developed rapidly in recent years, expanding the capabilities of LLMs to speech-related tasks through pre-trained speech encoders~\cite{AudioGPT, SpeechGPT}. When dealing with Automatic Speech Recognition (ASR) tasks, for example, prevailing architectures of LLM-based ASR typically employ a pre-trained speech encoder to transform acoustic signals into feature representations, which are then injected into LLMs as prompts for end-to-end ASR training~\cite{LLM-ASR-SE, seed-ASR}. Furthermore, LALMs architecture holds the potential to extend capabilities to more complex acoustic environments, such as Target-Speaker ASR (TS-ASR) in cocktail party scenarios, which remains a challenging research frontier.


Speech recognition in multi-talker scenario aims to transcribe speech signals containing overlapping utterances from multiple speakers into text. Current tasks in this scenario mainly fall into two categories: Multi-Talker ASR (MT-ASR) and Target-Speaker ASR (TS-ASR). MT-ASR~\cite{MT-ASR-25,MT-ASR-LLM} seeks to transcribe all speakers present in a mixed audio signal. The dominant approach is Serialized Output Training (SOT)\cite{SOT} which outputs transcriptions sequentially, typically separated by a special token, following a “first-in, first-out” strategy. While this simplifies the recognition process, it does not associate each transcription with a specific speaker, making it more suitable for applications like meeting transcription.

\begin{figure*}[t]
\begin{center}
\includegraphics[width=0.9\linewidth]{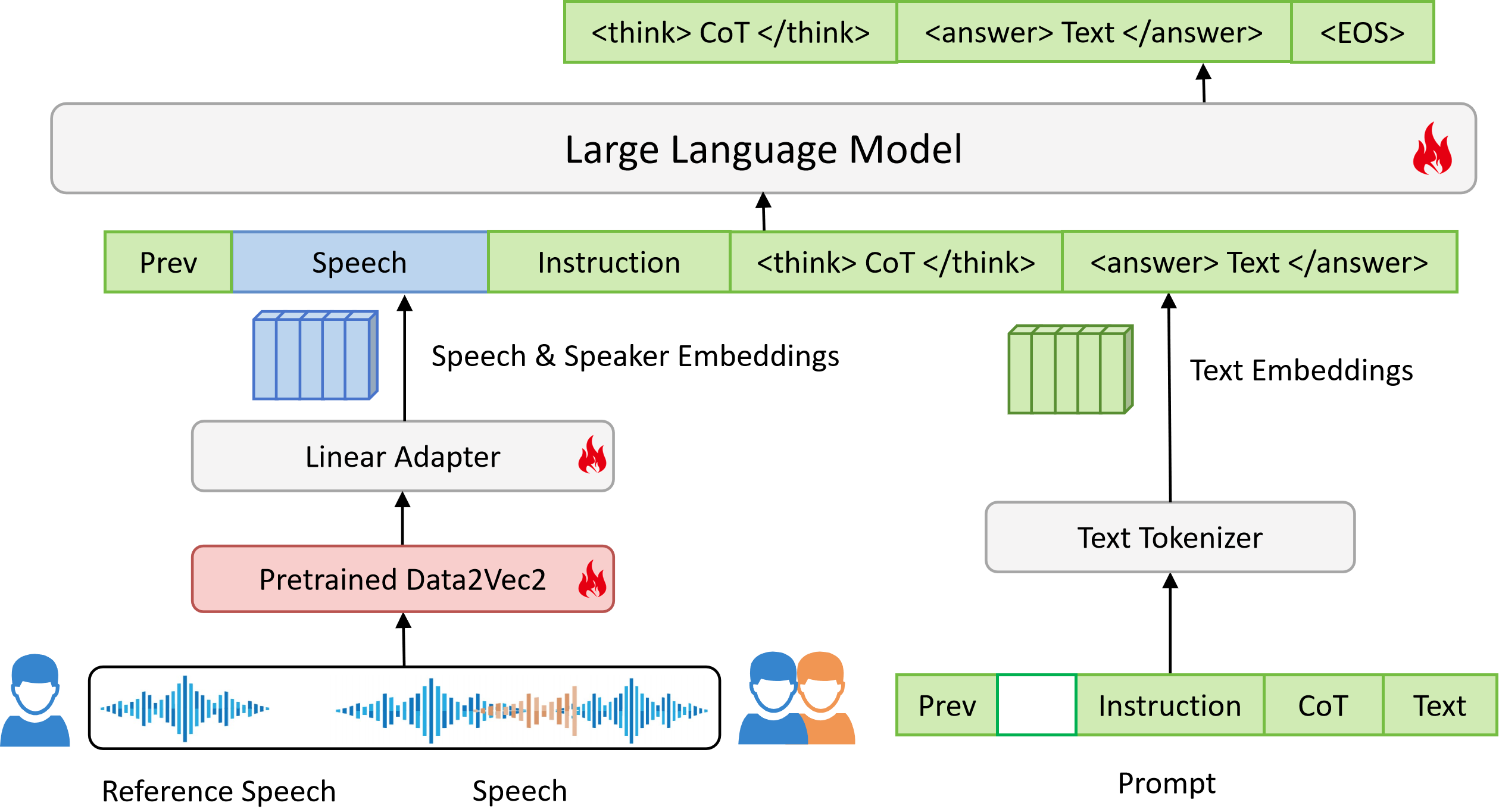}
\end{center}
\caption{The overview of TS-ASR framework based on LALM architecture.}
\label{fig:Overview}
\end{figure*}

In contrast, many applications such as smart home assistants, and conversational agents require accurately recognizing speech of a specific user in noisy, multi-speaker environments. To address these requirements, TS-ASR has been widely studied. Given a reference speech of the target speaker, TS-ASR selectively transcribes the speech of the target speaker while suppressing interference from others~\cite{TS-ASR-2019}. The TS-ASR task relies on additional reference information and is sensitive to speaker verification errors, making it a challenge task in multi-talker scenario. Some previous works designed a neural network architecture to jointly train Target Speaker Extraction (TSE) and ASR in order to transcribe the speech of the target speaker~\cite{TS-ASR-conformer, TS-ASR-WavLM}. There were also some works integrating speaker embeddings into the training of conventional end-to-end ASR models to directly train the TS-ASR model~\cite{SQ-Whisper, TS-ASR-Whisper, TS-VAD}. With the recent advancement of LALMs, MT-LLM~\cite{MT-LLM}, which utilized both an ASR encoder and a speaker encoder to extract features and then fed them into the LLM backbone to get TS-ASR output, has also demonstrated a good performance on the TS-ASR task. Although the development of LALMs has already brought new insights into TS-ASR, the optimization potential remains far from exhausted, as TS-ASR is a complex task that requires deep audio comprehension to achieve higher accuracy.

The breakthrough of DeepSeek-R1~\cite{DeepSeek-Nature} shows that combining CoT with RL significantly improves reasoning in complex tasks. This approach has also proven to be effective on many speech tasks, such as Audio Question and Answering (AQA)~\cite{Omni-R1, Audio-CoT} and ASR~\cite{GRPO-ASR}. TS-ASR is a complex task that requires models to deeply understand speech and distinguish between different speakers, which involves logical reasoning to determine the number of speakers, identify the target speaker, and select the relevant speech segments for recognition. Therefore, we believe that CoT and RL are able to improve the performance of TS-ASR task by guiding the model to produce explicit intermediate reasoning steps before arriving at the final answer.

In this work, we design a set of CoT and RL training method adapted for the TS-ASR task to enhance reasoning capabilities in cocktail party scenarios. First, a TS-ASR \textit{BASE} model is trained based on the LALMs architecture. Then, a novel method for constructing CoT training data is proposed, in which key information such as speaker count, overlap duration, speaker gender, and similarity to the reference is extracted from mixed speech. These attributes are subsequently structured through logical organization to form reasoning data for the CoT training, which is fine-tuned on the TS-ASR \textit{BASE} model. Finally, based on the model trained on CoT data, RL training is further conducted with selected data to enhance the generalization capabilities of the model. Experiment results show a significant improvement of TS-ASR performance with proposed CoT training method, and a further improvement is achieved after conducting the proposed RL training method. This demonstrates the effectiveness of the proposed CoT and RL training method designed for the TS-ASR task.


\begin{figure*}[t]
\begin{center}
\includegraphics[width=1\linewidth]{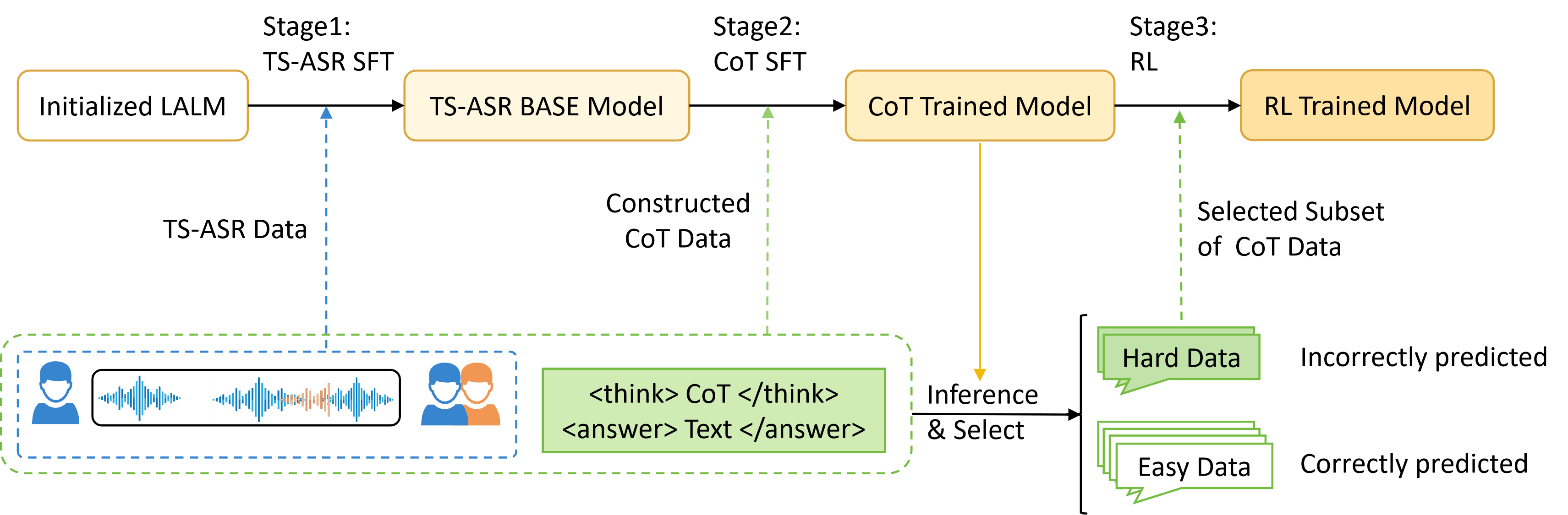}
\end{center}
\caption{Training pipeline of the proposed method.}
\label{fig:Pileline}
\end{figure*}

In summary, our contributions are as follows: 
\begin{itemize}
\item This is the first work to propose that logical reasoning benefits the TS-ASR task, and a set of CoT and RL training methods adapted for the TS-ASR task are thereby proposed.

\item We construct a novel CoT training dataset for TS-ASR by extracting task-relevant information in logic from the mixed speech. This CoT dataset will be open-sourced to support further research at https://anonymous.4open.science/r/TS-ASR-CoT-RL-anonymous-1DFF
\item Our approach shows a significant improvement of TS-ASR performance with CoT and RL training, which demonstrates the effectiveness of the proposed CoT and RL training methods adapted for the TS-ASR task, establishing a novel framework that enhances reasoning capabilities in cocktail party scenarios.
\end{itemize}

\section{Methodology}

\label{sec:method}
\subsection{LALM-based TS-ASR}

\subsubsection{Model structure}

The proposed framework of the TS-ASR model is shown in Figure~\ref{fig:Overview}. It consists of speech encoder, linear adapter, text tokenizer, and an LLM backbone network. We select a 3s segment from the target speaker utterance as reference, concatenate it with the mixed input (separated by a 3s silence), and feed the concatenated audio into the speech encoder. The pre-trained speech encoder Data2Vec2~\cite{d2v2} converts the input speech into frame-level embedding vectors that contain semantic and speaker information. To achieve modality alignment, a linear adapter maps the output from the speech encoder to the feature space of LLM. Meanwhile, the text tokenizer transforms the textual prompt into text embeddings. The speech and text embeddings are then combined and fed into the LLM. After fine-tuning, the model can focus on the target speaker within the mixed speech and generate accurate transcriptions. 

In our framework, reference speech is concatenated with mixed speech as an audio prompt~\cite{Time-ref}. This approach effectively leverages the inherent capabilities of LLMs in generating targeted responses from prompts for TS-ASR. Moreover, the architecture employs only the Data2Vec2 speech encoder without introducing a separate speaker encoder. As speaker and semantic information are both inherently captured by the Self-Supervised Learning (SSL) model~\cite{wavlm}, their fusion is naturally achieved within the encoder.

\subsubsection{Training pipeline}

Our TS-ASR training follows a three-stage paradigm as shown in Figure~\ref{fig:Pileline}:

\textbf{TS-ASR \textit{BASE}} A base TS-ASR model is trained via full parameter Supervised Fine-Tuning (SFT) on the initialized LALM, where all components are jointly updated, including the pre-trained LLM, the pre-trained speech encoder, and the randomly initialized adapter. This step aims to align multi-modal information and enable the model to perform TS-ASR. This step leverages TS-ASR data that pairs multi-talker speech with target speaker references, effectively addresses the cold start problem and provides a foundation for subsequent CoT training. 

\begin{figure*}[t]
\begin{center}
\includegraphics[width=1\linewidth]{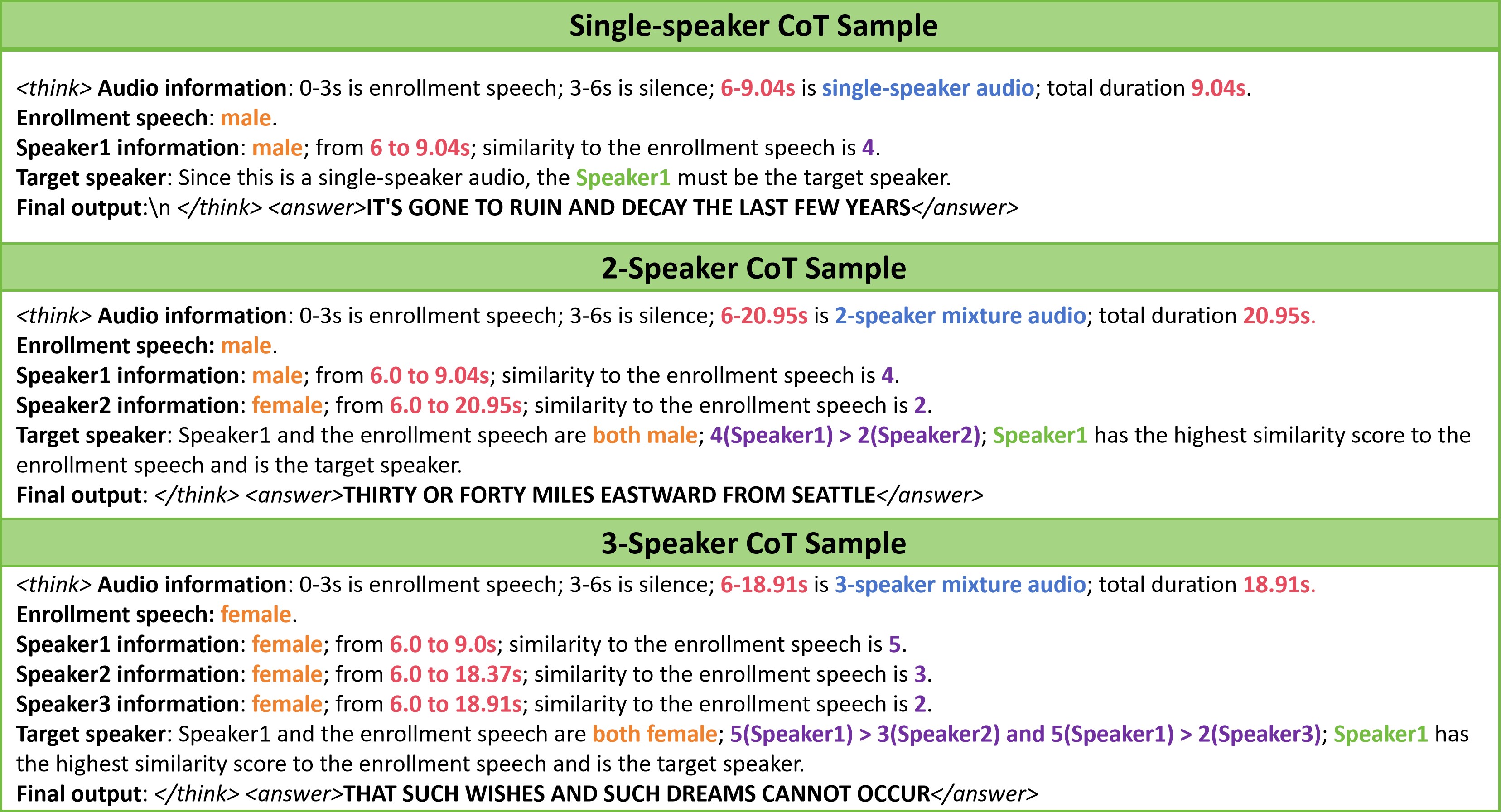}
\end{center}
\caption{Sample of CoT data for single-speaker, 2-speaker and 3-speaker mixed speech. Text in different colors represents different types of information.}
\label{fig:CotData}
\end{figure*}

\textbf{CoT Training} 
The TS-ASR \textit{BASE} model is fine-tuned with constructed CoT data to guide step-by-step reasoning. This teaches the model to first analyze the input speech by describing the reference speech and the attributes of each speaker, then reason step by step to identify the target speaker, and finally generate the transcription. The core idea is that, for complex tasks, LLMs often struggle to produce correct answers directly. The CoT method addresses this limitation by explicitly providing reasoning steps, guiding the model to think step by step like humans and thereby improving its reasoning capability.

\textbf{RL Training} Furthermore, based on the CoT trained model, RL refines the reasoning steps and improves the final performance using Group Relative Policy Optimization (GRPO)~\cite{Deepseekmath} with Word Error Rate (WER) and CoT format rewards. In order to further focus learning on challenging cases, the RL training set is constructed by selecting a "Hard Data" subset of the original CoT data, which consists of examples where the CoT-trained model predicted either incorrectly CoT format or erroneous ASR transcriptions. 



\subsection{CoT Training}

\subsubsection{CoT data construction}
Our CoT dataset is constructed based on LibriSpeech~\cite{LibriSpeech} dataset, which contains gender information, speaker information, and ASR transcription for each single utterance. 
As shown in Figure~\ref{fig:CotData}, since the reasoning structure required for TS-ASR is relatively fixed, a standardized data template is designed that includes five components. First, audio information describes the composition of the input speech, the number of speakers, and the total duration. Second, reference speech primarily records the gender of the target speaker. Third, the corresponding speech of each speaker is described by timestamps, gender, and a similarity level to the reference. Since the mixed speech is generated from single-speaker utterances, these information can be obtained from the source single-speaker speech. Fourth, the target speaker is then inferred based on the above information. In single-speaker scenarios, identification is straightforward, while in multi-speaker scenarios, gender and similarity level are jointly considered to reach a decision. Finally, the answer is the transcription corresponding to the target speaker. 

To compute similarity levels, we use CAM++~\cite{cam++} to extract speaker embeddings and calculate the cosine similarity scores between embeddings of each source speech and the reference speech. Then, these continuous scores in $[-1,1]$ are mapped to five discrete levels $[1,5]$, defined as:
\begin{equation}
    l =
    \begin{cases}
        1, & \text{if } s < 0, \\
        1 + \lfloor 5 \times s \rfloor, & \text{if } 0 \le s < 1, \\
        5, & \text{if } s = 1.
    \end{cases}
\end{equation}
where $s$ denotes the similarity score and $l$ denotes the similarity level. Specifically, the  continuous score $s$ is multiplied by 5, rounded down to the nearest integer, and then shifted by an offset of 1 to map the result into the range from 1 to 5. For the few cases where $s<0$, the score is clamped to the minimum level of 1. This discretization prevents the model from focusing on insignificant numerical differences, thereby improving training stability and convergence efficiency.

\subsubsection{Training with CoT data}
\label{sec:cot_train}

Since LLMs often struggle with complex tasks, CoT guides them with explicit reasoning steps like humans~\cite{CoT-complextask}. In this stage, the model is trained to analyze the speech content, estimate the number of speakers and describe the attributes of each speaker. Based on the structured analysis, the model reasons step by step to determine which speaker is the target. Finally, the model generates transcription from the identified target speaker. This staged reasoning not only improves model performance but also enhances the interpretability and reliability of its predictions.



\subsection{RL Training}

\subsubsection{Data selection}

High quality data benefits the RL training\cite{RL-audio}, therefore a data selection strategy is designed for the RL training of TS-ASR, focusing on samples most beneficial for performance improvement. As shown in Figure~\ref{fig:Pileline}, inference is performed on the full training set using the model trained by CoT data, selecting all samples with incorrectly predicted CoT format and a random subset of samples with correct format but wrong ASR results. ASR errors serve as a direct indicator of task difficulty, while format errors are fundamental but important for format maintaining. Including both types of error data in RL training helps the model further refine its reasoning capabilities within the specified format. With this data selection strategy, RL training can focus on challenging samples, improving both discrimination and generalization capabilities on challenging cases.






\subsubsection{Group Relative Policy Optimization}
GRPO algorithm is adopted to train the model. For each input, a group of generated outputs are sampled and ranked within the group based on task-specific rewards. This approach effectively combines positive rewards that reinforce correct outputs and negative rewards that penalize uncertain or incorrect predictions. 

In GRPO, for each input $x$, a group of $G$ outputs is generated and compared within the group based on task-specific rewards $r_k$. The normalized advantage $\hat{A}_k$ of the $k$-th response $y_k$ is:
\begin{align}
    \hat{A}_k = \frac{r_{k} - \mathrm{mean}({\{r_{1},...,r_{G}\}})}{\mathrm{std}(\{r_{1},...,r_{G}\})}
\end{align}
where $r_k$ is the reward of the response $k$.

The final GRPO objective is defined as:
\begin{equation}
\begin{split}
\mathcal{J}_{\mathrm{GRPO}}(\theta) = \frac{1}{G} \sum_{k=1}^{G} \frac{1}{|\boldsymbol{y}_k|} \sum_{t=1}^{|\boldsymbol{y}_k|} 
& \min \big\{ \rho_{k}(\theta) \cdot \hat{A}_k, \\
& \mathrm{clip}\big(\rho_{k}(\theta), 1-\epsilon, 1+\epsilon\big) \cdot \hat{A}_k \big\}
\end{split}
\end{equation}
where the clip function limits the relative probability ratio $\rho_{k}(\theta)$ to a range defined by the hyperparameter $\epsilon$. 

\begin{equation}
\rho_{k}(\theta) = \frac{\pi_{\theta}(y_{k} \mid x)}{\pi_{\theta_{\mathrm{old}}}(y_{k} \mid x)}
\end{equation}

Here, $\pi_{\theta}$ and $\pi_{\theta_{\mathrm{old}}}$ are the current and old policy models, respectively.

\subsubsection{GRPO for TS-ASR}

In TS-ASR, the reward $r$ for GRPO training is the combination of WER reward $r_{\text{WER}}$ and CoT format reward $r_{\text{format}}$:
\begin{equation}
r = r_{\text{WER}} + r_{\text{format}}.
\label{eq:reward}
\end{equation}
The WER reward evaluates the accuracy of the transcription between \textless answer\textgreater and \textless /answer\textgreater.
Sequences with a lower WER are assigned higher rewards, offering a direct guidance for speech recognition, defined as:
\begin{equation}
r_{\text{WER}} = 1-\frac{Sub+Del+Ins}{N},
\label{eq:reward_wer}
\end{equation}
where $Sub,Del,Ins$ are substitution, deletion and insertion errors, respectively. And $N$ is the number of words in ground-truth. The format reward is:
\begin{equation}
r_{\text{format}} = 
\begin{cases}
1 & \text{if  } y \in F, \\
0 & otherwise,
\end{cases}
\label{eq:reward_format}
\end{equation}
where $F$ is the format: $<think>...</think><answer>...</answer>$ and $y$ is the output string. The format reward is designed to address cases where the model occasionally incompletely generates the required \textless think\textgreater and \textless answer\textgreater tags. It checks the completeness of these structural elements and encourages the model to perform the full reasoning process and ensures correct reasoning structure. 


\section{Experiment}
\label{sec:pagestyle}
\subsection{Dataset}

Based on the approach in LibriMix~\cite{librimix}, this study generates mixed speech datasets of two speakers (Libri2Mix) and three speakers (Libri3Mix) using the 960-hour LibriSpeech~\cite{LibriSpeech} single-speaker corpus\footnote{https://github.com/JorisCos/LibriMix}. Following the data scale in ~\cite{MT-LLM}, we create mixed samples for each source utterance, alternately selecting each speaker in the mixture as the target. Reference speech segments are randomly selected from the single utterances of the same speaker. The duration of the Libri2Mix and Libri3Mix datasets is 1920 and 2880 hours, respectively. When combined with the original 960 hours of single-speaker data, the total training set reaches about 5760 hours. The test set is generated from LibriSpeech test-clean, with Libri2Mix and Libri3Mix constructed using the mixture and reference speech provided by the open-source SpeakerBeam toolkit\footnote{https://github.com/BUTSpeechFIT/speakerbeam/tree/main/egs}.

\subsection{Experiment Settings}
The proposed model uses Qwen2.5-0.5B-Instruct~\cite{qwen2.5} as the backbone LLM. It adopts a pre-trained Data2Vec2 model as the audio encoder, which is trained on one million hours of unsupervised speech data. The number of parameters of Data2Vec2 is 0.3 billion. 

During the SFT stage, the AdamW optimizer~\cite{AdamW} is applied with a learning rate schedule combining linear warmup and cosine annealing. The learning rate is gradually increased from 1e-7 to 7e-5 over the first 3000 steps and decayed to a minimum of 1e-6. The model is trained for 240000 steps, equivalent to one full epoch, with a global batch size of 112. To prevent input truncation, the maximum text length is set to 512. Finally, the model with the best performance on the validation set is selected as the final model.

For RL training, the model is fine-tuned on 20000 selected samples for 2000 steps with a learning rate of 1e-6 and a temperature of 1. For each input, 8 responses are generated, with a maximum length of 512 tokens. And the model from the final training step is selected as the final model.

\subsection{Evaluation Metrics}
In this experiment, the evaluation metric is WER, which is calculated by extracting the text between \textless answer\textgreater and \textless /answer\textgreater tags from the decoded output of the LibriSpeech, Libri2Mix, and Libri3Mix test sets. For the small number of outputs with formatting errors or incomplete tags, the result is treated as empty for evaluation, resulting in deletion errors. The reported WER is computed only for the target speaker. In multi-speaker mixtures, each speaker is independently treated as the target, and the WER is evaluated separately for each speaker.


\section{Result and Discussion}
\label{sec:typestyle}

\subsection{Comparison to Baselines}
We compare our proposed model with both traditional TS-ASR methods and LLM-based approaches. As shown in Table~\ref{tab:Baseline}, our model achieves significant improvements, with WERs of 4.84\% on Libri2Mix and 12.23\% on Libri3Mix.

\begin{table}[ht]
\caption{TS-ASR performance on Libri2Mix and Libri3Mix test-clean dataset. Evaluated by WER(\%).}\label{tab:Baseline}
\setlength{\tabcolsep}{6pt}
\centering 
\begin{tabular}{lcc}
\hline
\multirow{2}{*}{\textbf{Model}}   & \multicolumn{2}{c}{\textbf{LibriMix}}\\
\cline{2-3}
 & \textbf{2-mix} & \textbf{3-mix} \\
\cline{1-3}
 Whisper + LoRA~\cite{TS-ASR-Whisper} & 11.98 & - \\
 MUSE-TSASR ~\cite{MUSE-TSASR} & 10.53 & - \\
 WhisperTSE-L ~\cite{WhisperTSE-L} & 8.1 & - \\
 Whisper-SS-TTI~\cite{Whisper-MT+TS-ASR} & 7.97 & 21.97 \\
 WavLM + TSE~\cite{TS-ASR-WavLM}  & 7.6 & - \\ 
 TS-VAD(460h) ~\cite{TS-VAD}  & 6.61 & 14.81 \\ 
 MT-LLM ~\cite{MT-LLM}& 6.7 & 16.2 \\ 
\hline
 TS-ASR Base  & 7.4 & 17.24 \\
 TS-ASR + CoT & 5.29 & 13.57 \\
 TS-ASR + CoT + RL & \textbf{4.84} & \textbf{12.23} \\
\hline
\end{tabular}
\end{table}

Compared with MT-LLM, the current best-performing LLM-based TS-ASR model, our TS-ASR \textit{BASE} model without CoT data has a lower performance. However, training with CoT data achieves a significant improvement, reducing the WER on Libri2Mix and Libri3Mix by 21\% and 16.2\%, respectively. At this stage, the proposed model already outperforms previous work on the LibriMix dataset. It demonstrates that logic reasoning helps the model more accurately transcribe the target speaker.


Based on the model trained on CoT data, RL fine-tuning is applied to selected data, further reducing the WER of 2-mix and 3-mix to 4.84\% and 12.23\%, respectively. It indicates that RL training can help the model further improve the reasoning capabilities. Specifically, we observe that the rate of incomplete \textless think\textgreater and \textless answer\textgreater tags decreases from 0.4\textperthousand \, to 0.23\textperthousand \, after RL fine-tuning, indicating that the format reward further ensures the completeness of the reasoning structure.
In terms of final performance, our model achieves substantially lower WER than all baseline models, achieving new SOTA performance in Libri2Mix and Libri3Mix dataset. This confirms that our exploration of LLM-based CoT and RL for TS-ASR represents a valuable and effective direction for research.




\begin{table}[ht]
\caption{Comparison of similarity scores and similarity levels for CoT data, evaluated by WER(\%).}\label{tab:CoT}
\setlength{\tabcolsep}{14pt}
\centering 
\begin{tabular}{lcc}
\hline
\multirow{2}{*}{\textbf{Model}}   & \multicolumn{2}{c}{\textbf{LibriMix}}\\
\cline{2-3}
 & \textbf{2-mix} & \textbf{3-mix}\\
\cline{1-3}
 TS-ASR + CoT \\
\hspace{8pt} + similarity score  & 5.39 & 14.09  \\
\hspace{8pt} + similarity level  & 5.29 & 13.57 \\
\hline
\end{tabular}
\end{table}

\subsection{Effectiveness of Discrete Similarity Levels}

We explore the impact of continuous similarity scores and discrete similarity levels on model performance to identify an effective configuration. As shown in Table~\ref{tab:CoT}, the model trained on the data using similarity scores is compared to that trained on data with discrete similarity levels. Results show that using discrete similarity levels reduces WER by 0.1\% and 0.52\%, with consistent improvements observed on both Libri2Mix and Libri3Mix. 

The improvement shows the effectiveness of replacing continuous floating-point scores with discrete levels in the CoT data. This further indicates that detailed floating values are more difficult to understand and leverage by LLMs than concise discrete similarity levels.





\subsection{Ablation Studies for CoT Template}\label{sec:CoT-tplt}

\begin{figure}[t]
\begin{center}
\includegraphics[width=1\linewidth]{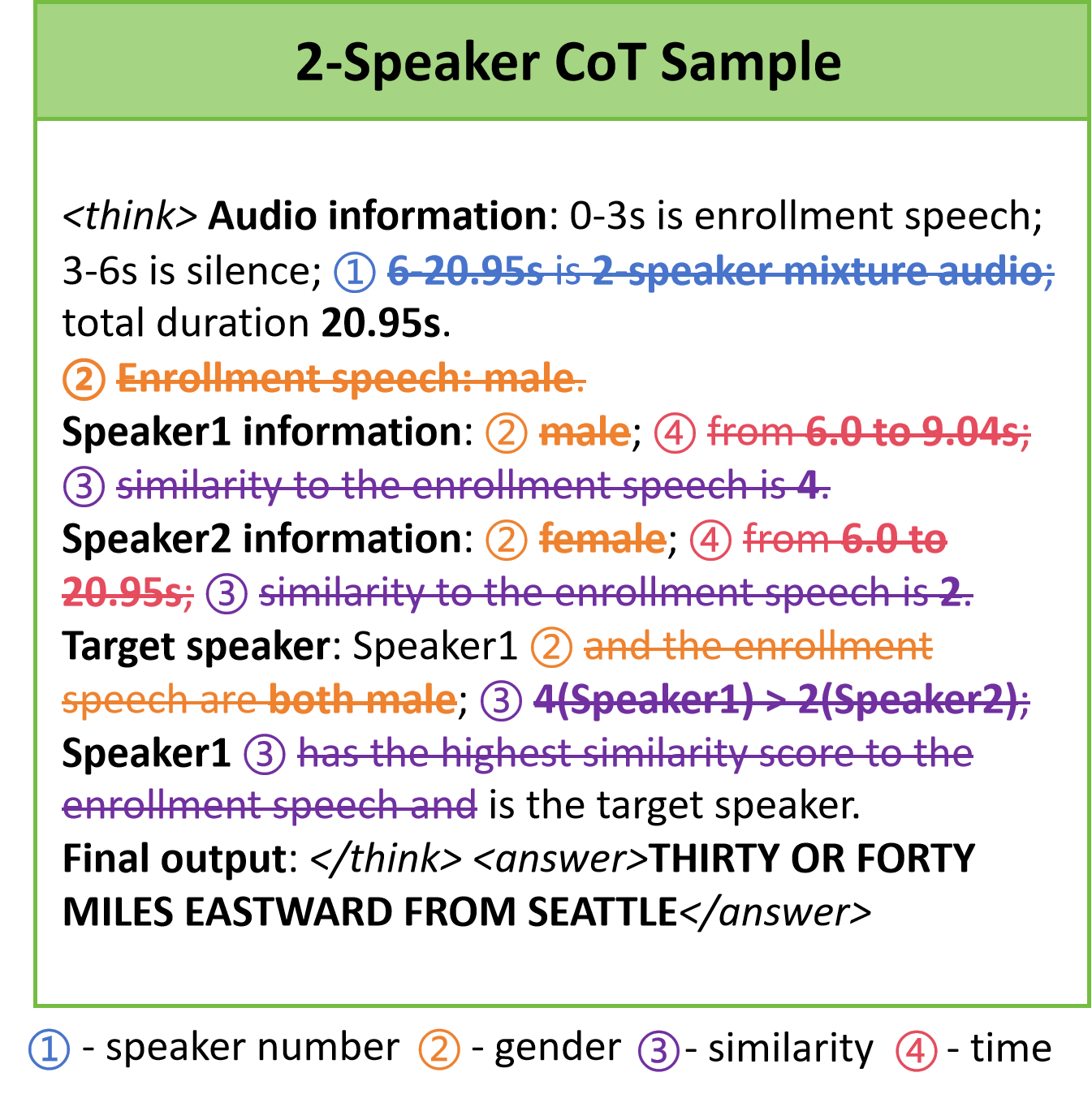}
\end{center}
\caption{A sample of CoT data for 2-speaker mixed speech. Different types of removed information are being marked with distinct colors and numbers in the text.}
\label{fig:CotDataDel}
\end{figure}

This section presents ablation studies on the key variables in the CoT template to evaluate the importance of different types of information for TS-ASR reasoning. As shown in Figure~\ref{fig:CotDataDel}, the CoT data are processed by individually removing information related to the number of speakers, gender, speaker similarity, and start and end times, followed by retraining the CoT model based on the TS-ASR BASE model.

\begin{table}[ht]
\vspace*{5pt}
\caption{Ablation Studies for CoT template, evaluated by WER(\%). "-" means the removal of specific information from the full CoT data.}\label{tab:CoT-tplt}
\setlength{\tabcolsep}{14pt}
\centering 
\begin{tabular}{lcc}
\hline
\multirow{2}{*}{\textbf{Model}}   & \multicolumn{2}{c}{\textbf{LibriMix}}\\
\cline{2-3}
 & \textbf{2-mix} & \textbf{3-mix}\\
\cline{1-3}
TS-ASR + full CoT data & 5.29 & 13.57  \\
\hspace{8pt} - speaker number  & 5.46 & 14.01 \\
\hspace{8pt} - speaker gender  & 5.32 & 13.7 \\
\hspace{8pt} - speaker similarity  & 6.08 & 14.93 \\
\hspace{8pt} - start and end times  & 5.19 & 13.54 \\
\hline
\end{tabular}
\end{table}

\vspace*{5pt}
In Table~\ref{tab:CoT-tplt}, the results show that removing speaker similarity leads to a significant performance drop. The WER increases from 5.29\% to 6.08\% on Libri2Mix and from 13.57\% to 14.93\% on Libri3Mix. This indicates that speaker similarity is the most critical feature for CoT reasoning. Removing the number of speakers also causes a noticeable increase in WER, particularly in the 3-speaker mixture, where WER rises by 0.43\%, suggesting that this information is also important. Meanwhile, removing either gender or time information affects performance by no more than 0.2\% across all test sets, with no significant difference observed.



This conclusion aligns with human auditory perception. In noisy environments, listeners rely on speaker similarity to identify the target speaker. At the same time, the number of speakers, which reflects the complexity of the acoustic environment, affects how clearly the target speech can be heard. Together, these two factors are essential for speech understanding in cocktail party scenarios.


\begin{figure}[t]
\begin{center}
\includegraphics[width=1\linewidth]{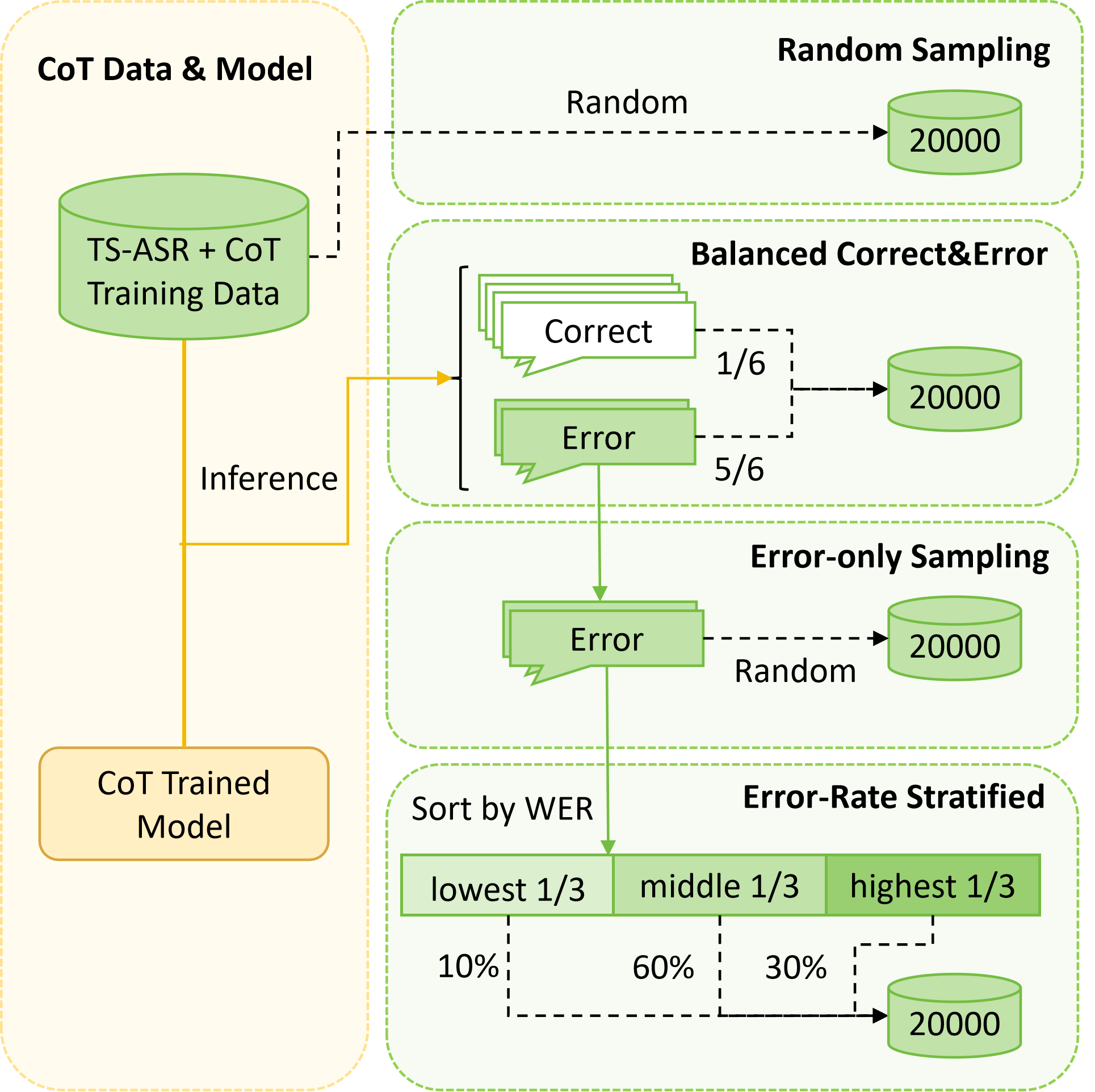}
\end{center}
\caption{Schematic illustration of different RL data selection strategies. All strategies select 20,000 samples as RL training data.}
\label{fig:RL-Selection}
\end{figure}

\subsection{Ablation Studies for RL Data Selection}
Since data selection is important for effective RL training, experiments are conducted on different data selection methods for the RL training of TS-ASR task. Table~\ref{tab:RL} presents the impact of different RL training data selection strategies on model performance. As shown in Figure~\ref{fig:RL-Selection}, “Random Sampling” denotes randomly selecting a subset from the full training dataset for RL training. “Balanced Correct\&Error” selects correct and error samples in a 1:5 ratio from the model trained on CoT data for RL training, aiming to focus on difficult cases while preserving general performance. The "Error-only Sampling" strategy uses only the subset of incorrectly predicted samples for subsequent training. “Error-Rate Stratified” sorts all error samples in ascending order of WER and then randomly selects a total of 20,000 samples from the lowest 1/3, middle 1/3, and highest 1/3, with the number of samples drawn from each group following a ratio of 1:6:3.

\begin{table}[ht]
\caption{Ablation Studies for RL data selection, evaluated by WER(\%).}\label{tab:RL}
\setlength{\tabcolsep}{11pt}
\centering 
\begin{tabular}{lcc}
\hline
\multirow{2}{*}{\textbf{Model}}   & \multicolumn{2}{c}{\textbf{LibriMix}}\\
\cline{2-3}
 & \textbf{2-mix} & \textbf{3-mix} \\
\cline{1-3}
\multicolumn{3}{l}{TS-ASR + CoT (with similarity level) + RL} \\
\hspace{8pt} + Random Sampling  & 4.86 & 12.72  \\
\hspace{8pt} + Balanced Correct\&Error  & 4.85 & 12.43  \\
\hspace{8pt} + Error-Rate Stratified  & \textbf{4.8} & 12.34  \\
\hspace{8pt} + Error-only Sampling & 4.84 & \textbf{12.23} \\
\hline
\multicolumn{3}{l}{TS-ASR + CoT (with similarity level - time) + RL} \\
\hspace{8pt} + Error-Rate Stratified  & 4.81 & 12.29 \\
\hspace{8pt} + Error-only Sampling & 4.84 & 12.32 \\
\hline
\end{tabular}
\end{table}

Experimental results show that the “Random Sampling” strategy performs the worst on Libri3Mix, with a WER of 12.72\%, while the “Error-only Sampling” strategy achieves the best result, reducing WER to 12.23\%. When considering Libri2Mix and Libri3Mix together, the "Error-Rate Stratified" strategy also performs well, achieving WERs of 4.80\% and 12.34\%, respectively. This approach also uses error only samples and further refines sample selection by stratifying them according to WER, thereby focusing more on the challenging cases within the error set. This demonstrates that in reinforcement learning for TS-ASR, challenging error samples contribute more to performance improvement than correct samples. Therefore, training with a higher proportion of error samples is more effective for helping the model correct mistakes and improve overall accuracy.

Since the CoT model without time information achieves marginal better performance in Table~\ref{tab:CoT-tplt}, we adopt this model and apply the two best data selection strategies for RL training. As shown in the last two rows of Table~\ref{tab:RL}, removing time information does not lead to better final performance compared to using the full CoT data. Therefore, the time information in the CoT data is retained.



\begin{table}[ht]
\caption{Ablation Studies for single speaker in LibriSpeech test-clean dataset, evaluated by WER(\%).}\label{tab:Sig}
\setlength{\tabcolsep}{16pt}
\centering 
\begin{tabular}{lc}
\hline
\textbf{Model}  & \textbf{LibriSpeech}\\
\cline{1-2}
TS-ASR Base  & 8.1 \\
TS-ASR + CoT  & 3.88   \\
TS-ASR + CoT + RL & 3.65  \\
\hline
\end{tabular}
\end{table}

\subsection{Comparison for Single-speaker}

MT-LLM \cite{MT-LLM} reports single-speaker ASR performance alongside its evaluation of multi-talker scenarios under the MT-ASR task. It is also important to report single-speaker performance for the TS-ASR task. In real-world applications, the system cannot know in advance whether the input is a multi-talker mixture or not. As a result, given a reference speech from the target speaker, a TS-ASR system should be required to transcribe an utterance containing only the speech of the target speaker. Consequently, it needs to handle both multi-talker mixtures and single-speaker scenarios. However, prior work has not paid enough attention to single-speaker TS-ASR performance on the LibriSpeech dataset.

To evaluate performance in this setting, ablation studies are conducted on the LibriSpeech test-clean dataset, with results shown in Table~\ref{tab:Sig}. Our evaluation follows the standard TS-ASR protocol by using the TS-ASR prompt and randomly selecting a reference utterance from the same speaker. The results show that our proposed improvements enhance TS-ASR performance not only in multi-speaker settings but also in the single-speaker case, with the model trained using CoT and RL achieving a WER of 3.65\% on the single-speaker test set.

Nevertheless, TS-ASR still faces unique challenges in single-speaker scenarios. It is observed that in single-speaker scenarios, the model tends to produce more deletion errors. This may be due to the fact that TS-ASR is designed to suppress non-target speakers in multi-talker environments, which can cause the suppression mechanism to mistakenly treat the target speech as interference in single-speaker inputs, leading to degraded performance.

\vspace*{3pt}
\section{Conclusion}
In this work, we proposed a novel framework that incorporates CoT and RL training into the TS-ASR task to enhance reasoning capabilities in cocktail party scenarios. The approach began with training a TS-ASR \textit{BASE} model using the LALM architecture. Subsequently, a novel methodology was proposed to generate CoT training samples by extracting salient features from overlapped speech, such as number of speakers, speech duration, gender information, and similarity to the reference speech. These features were systematically organized into structured reasoning data to fine-tune the TS-ASR \textit{BASE} model. Furthermore, a subset of samples from the CoT training phase was selected for additional RL-based refinement. Experimental results showed a significant improvement of TS-ASR performance with CoT and RL training, which not only demonstrated the effectiveness of our constructed CoT dataset and the proposed training framework but also established a SOTA performance in the Libri2Mix and Libri3Mix datasets on the TS-ASR task. The successful integration of CoT and RL into TS-ASR demonstrated the efficacy of logical reasoning in cocktail party scenarios.  As LALMs continue to advance with CoT and RL techniques, we believe this framework has significant potential for further solving the cocktail party problem.

\section*{Impact Statement}
This paper presents work whose goal is to advance the field of Machine
Learning. There are many potential societal consequences of our work, none
which we feel must be specifically highlighted here.


\bibliography{example_paper}
\bibliographystyle{icml2026}




\end{document}